\documentclass[twocolumn]{article}
\usepackage[utf8]{inputenc}
\usepackage{usenix}
\usepackage{amssymb}
\usepackage{amsmath}
\usepackage{booktabs}
\usepackage{nicefrac}
\usepackage{multirow}
\usepackage{censor}
\usepackage{rotating}
\usepackage{enumitem}
\usepackage{caption}
\usepackage{subcaption}
\usepackage{tabularx}
\usepackage{titlecaps}
\usepackage{array}
\usepackage{float}
\usepackage{stfloats}
\usepackage{fancyhdr}
\usepackage{fancyvrb}

\usepackage{xcolor}

\newcommand{\pii}{\censor{------}}

\newtheorem{definition}{Definition}
\DeclareMathOperator*{\argmax}{arg\,max}

\title{Extracting Training Data from Large Language Models}

\author{
\rm{Nicholas Carlini}$^1$ \and
\rm{Florian Tram\`er}$^2$ \and
\rm{Eric Wallace}$^3$ \and
\rm{Matthew Jagielski}$^4$ \and
\rm{Ariel Herbert}-Voss$^{5,6}$ \and
\rm{Katherine Lee}$^1$ \and
\rm{Adam Roberts}$^1$ \and
\rm{Tom Brown}$^5$ \and  \and
\rm{Dawn Song}$^3$ \and
\rm{\'{U}lfar Erlingsson}$^7$ \and
\rm{Alina Oprea}$^4$ \and
\rm{Colin Raffel}$^1$ \and
\mbox{$^1$Google\,\,\, 
$^2$Stanford\,\,\, 
$^3$UC Berkeley\,\,\,
$^4$Northeastern University\,\,\, 
$^5$OpenAI\,\,\, 
$^6$Harvard \,\,\, 
$^7$Apple  }
} %

\usepackage{graphicx}

\newcommand{\term}{eidetic memorization}

\begin{document}

\maketitle

\begin{abstract}
It has become common to publish large (billion parameter) language models
that have been trained on private datasets.
This paper demonstrates that in such settings, an adversary can
perform a \emph{training data extraction attack} to recover
individual training examples by querying the language model.

We demonstrate our attack on GPT-2,
a language model trained on scrapes of the public Internet, 
and are able to extract hundreds of verbatim text sequences 
from the model's training data.
These extracted examples include
(public) personally identifiable information (names, phone numbers, and email addresses),
IRC conversations,
code,
and 128-bit UUIDs.
Our attack is possible 
even though each of the above sequences are included
in just \emph{one} document in the training data.

We comprehensively evaluate
our extraction attack
to understand the factors that contribute to 
its success. Worryingly, we find that larger models
are more vulnerable than smaller models.
We conclude by drawing lessons
and discussing possible safeguards
for
training large language models.

\end{abstract}

\section{Introduction}

\begin{figure}[t]
    \centering
    \includegraphics{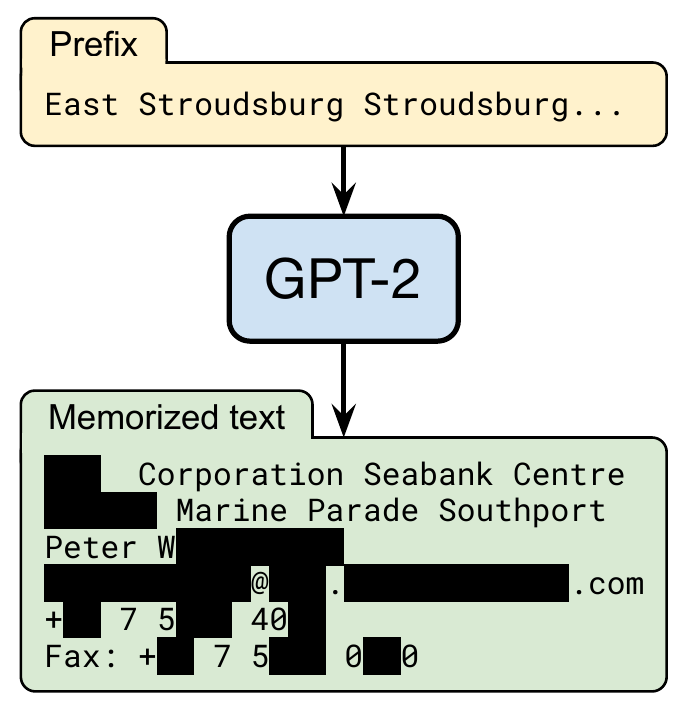}
    \vspace{-5pt}
    \caption{\textbf{Our extraction attack.} 
    Given query access to a neural network language model,
    we extract an individual person's name, email address, phone number, fax number, and physical
    address. The example in this figure shows information that is all accurate so we redact it to protect privacy.}
    \vspace{-5pt}
    \label{fig:teaser}
\end{figure}

Language models (LMs)---statistical models which assign a probability to a sequence of words---are fundamental to many natural language processing tasks.
Modern neural-network-based LMs use very large model architectures (e.g., 175 billion parameters~\cite{brown2020language}) and train on massive datasets (e.g., nearly a terabyte of English text~\cite{raffel2019exploring}). This scaling increases  the ability of LMs to generate fluent natural language~\cite{zhang2019dialogpt,radford2019better,zellers2019defending}, and also allows them to be applied to a plethora of other tasks~\cite{raffel2019exploring,hoang2019efficient,liu2019roberta}, even without updating their parameters~\cite{brown2020language}.

At the same time, machine learning models are notorious for exposing information about their (potentially private) training data---both in general \cite{shokri2017membership,nasr2019comprehensive} and in the specific case of language models~\cite{carlini2019secret,xkcd}. 
For instance, 
for certain models 
it is known that adversaries can apply \emph{membership inference attacks}~\cite{shokri2017membership}
to predict whether or not any particular example was in the training data.

Such privacy leakage is typically associated with \emph{overfitting} \cite{zhang2016understanding}---when a model's training error is significantly lower than its test error---because overfitting often indicates that a model has memorized examples from its training set. Indeed, overfitting is a sufficient condition for privacy leakage~\cite{yeom2018privacy} and many attacks work by exploiting overfitting~\cite{shokri2017membership}.

The association between overfitting and memorization
has---erroneously---led many to assume that 
state-of-the-art LMs will \emph{not} leak information about their training data.
Because these models
are often trained on massive de-duplicated datasets
only for a single epoch~\cite{brown2020language,raffel2019exploring},
they exhibit little to no overfitting~\cite{radford2019better}.
Accordingly,
the prevailing wisdom has been that
``the degree of
copying with respect to any given work is likely to be, at most, \emph{de minimis}'' \cite{eff}
and that models do not significantly memorize any particular training example.

\paragraph{Contributions.} In this work, we demonstrate that large language models memorize and leak individual training examples. 
In particular, we propose a simple and efficient method for extracting verbatim sequences from a language model's training set using only black-box query access.
Our key insight is that, although 
training examples do not have noticeably lower losses than test examples on \emph{average},
certain \emph{worst-case} training examples are indeed memorized. 

In our attack, we first generate a large, diverse set of high-likelihood samples from the model,
using one of three general-purpose sampling strategies. 
We then sort each sample using one of six different  metrics
that estimate the likelihood of each sample using a separate reference model (e.g., another LM), and rank highest the samples with an abnormally high likelihood ratio between the two models.

Our attacks directly apply to any language model, including those trained on sensitive and non-public data \cite{smartcompose,esteva2017dermatologist}. 
We use the GPT-2 model~\cite{radford2019language} released by OpenAI as a representative language model
in our experiments.
We choose to attack GPT-2 to minimize real-world harm---the GPT-2 model and original training data source are already public.

To make our results quantitative, we define a testable definition of memorization.
We then generate $1{,}800$ candidate memorized samples, $100$ under each of the $3 \times 6$
attack configurations, and find that over $600$ of them are verbatim samples from the GPT-2 training data (confirmed
in collaboration with the creators of GPT-2). %
In the best attack configuration, 67\% of candidate samples are verbatim training examples.
Our most obviously-sensitive attack extracts the full name, physical address, email address, phone number, and fax number of an individual (see Figure~\ref{fig:teaser}).
We comprehensively analyze our attack, including studying how model size and string frequency affects memorization, as well as how different attack configurations change the types of extracted data.

We conclude by discussing numerous practical strategies to mitigate privacy leakage. For example, differentially-private training \cite{abadi2016deep} is theoretically well-founded and guaranteed to produce private models if applied at an appropriate record level, but it can result in longer training times and typically degrades utility. We also make recommendations, such as carefully de-duplicating documents, that empirically will help to mitigate memorization but cannot prevent all attacks.

\iffalse
\eric{It would be good to mention some of the following at some point in the intro:
\begin{enumerate}
	\item differentially-private training is a principled defense against this type of privacy leakage. However, getting complete protection using DP causes unsatisfactory degradations in LM performance.  %
	\item We propose a novel, concrete definition of what memorization means for text. %
	\item We show that larger models memorize more, which means that this problem will get worse over time.
	\item we analyze which types of text are more likely to be memorized. %
	\item We propose practical suggestions for how to collect data and train models in order to minimize memorization. %
\end{enumerate}}
\fi

\section{Background \& Related Work}

To begin, we introduce the relevant background on large (billion-parameter) neural network-based
language models (LMs) as well as data privacy attacks.

\subsection{Language Modeling}\label{subsec:lm}

Language models are a fundamental building block of current
state-of-the-art natural language processing pipelines \cite{radford2018improving,devlin2018bert,howard2018universal,peters2018deep,raffel2019exploring}.
While the unsupervised objectives used to train these models vary,
one popular choice is a ``next-step prediction'' objective~\cite{bengio2003neural,mikolov2010recurrent,howard2018universal,radford2018improving}.
This approach constructs a generative model of the distribution
\setlength{\abovedisplayskip}{4pt}
\setlength{\belowdisplayskip}{4pt}
\[\textbf{Pr}(x_1, x_2, \ldots, x_n),\]
where $x_1, x_2, \ldots, x_n$ is a sequence of tokens from a vocabulary $\mathcal{V}$
by applying the chain rule of probability
\[\textbf{Pr}(x_1, x_2, \dots, x_n) = \Pi_{i=1}^n \textbf{Pr}(x_i \mid x_1, \dots, x_{i-1}).\]

State-of-the-art LMs use neural networks to estimate this probability distribution.
We let $f_\theta(x_i \mid x_1, \dots, x_{i-1})$ denote the likelihood of
token $x_i$ when evaluating the neural network
$f$ with parameters $\theta$.
While recurrent neural networks (RNNs) \cite{mikolov2010recurrent,graves2013generating} used to be a common choice for the neural network architecture of LMs,
\emph{attention}-based models \cite{bahdanau2014neural} have 
recently replaced RNNs in state-of-the-art models. In particular, 
\emph{Transformer} LMs \cite{vaswani2017attention} consist of a sequence of
attention layers and are the current model architecture of choice.
Because we believe our results are independent of the exact architecture used, we will not describe the Transformer architecture
in detail here and instead refer to existing work
\cite{alammar2018illustrated}.

\paragraph{Training Objective.} 
A language model is trained to maximize the probability of the data in a training set $\mathcal{X}$.
In this paper, each training example is a text document---for example, a specific news article or webpage from the internet.
Formally, training involves minimizing the loss function
\[\mathcal{L}(\theta) = -\log \Pi_{i=1}^n f_\theta(x_i \mid x_1, \dots, x_{i-1})\]
over each training example in the training dataset $\mathcal{X}$.
Because of this training setup, the ``optimal'' solution to the task of language modeling is to \emph{memorize} the answer to the question ``what token follows the 
sequence $x_1, \dots, x_{i-1}$?'' for every prefix in the training set.
However, state-of-the-art LMs are trained with massive datasets, which causes them to not exhibit significant forms of memorization: empirically, the training loss and the test loss are nearly identical \cite{radford2019better,raffel2019exploring,brown2020language}.

\paragraph{Generating Text.} 
A language model can generate new text (potentially conditioned
on some prefix $x_1, \dots, x_i$) by iteratively sampling
$\hat{x}_{i+1} \sim f_\theta(x_{i+1} | x_1, \dots, x_i)$
and then feeding $\hat{x}_{i+1}$ back into the model to sample
$\hat{x}_{i+2} \sim f_\theta(x_{i+2} | x_1, \dots, \hat{x}_{i+1})$.
This process is repeated until a desired stopping criterion is reached.
Variations of this text generation method include deterministically choosing the
most-probable token rather than sampling (i.e., ``greedy'' sampling)
or setting all but
the top-$n$ probabilities to zero and renormalizing the probabilities before sampling (i.e., top-$n$ sampling\footnote{For notational clarity, we write top-$n$ instead of the more common top-$k$ because we will use the constant $k$ for a separate purpose.}~\cite{fan2018hierarchical}).

\paragraph{GPT-2.} 
Our paper focuses on the GPT variant of Transformer LMs \cite{radford2018improving,radford2019language,brown2020language}.
Specifically, we demonstrate our training data extraction attacks on GPT-2, a family of LMs that were all trained using the same dataset and training algorithm, but with varying model sizes.
GPT-2 uses a word-pieces \cite{sennrich2015neural} vocabulary with a byte pair encoder \cite{gage1994new}.

GPT-2 XL is the largest model with 1.5 billion parameters. For the remainder of this paper, the ``GPT-2'' model refers to this 1.5 billion parameter model or, when we specifically indicate this, its Small and Medium variants with 124 million and 334 million parameters, respectively.

The GPT-2 model family was trained on data scraped from the public Internet.
The authors collected a dataset by following outbound links from the social media website Reddit.
The webpages were cleaned of HTML, with only the document text retained, and then \emph{de-duplicated} at the document level.
This resulted in a final dataset of $40$GB of text data, 
over which the model was trained for approximately 12 epochs.\footnote{Personal communication with the GPT-2 authors.}
As a result, GPT-2 does not overfit: the training loss is only roughly $10\%$ smaller than the test loss across all model sizes.

\subsection{Training Data Privacy}

It is undesirable for models to remember any details that are specific to their (potentially private) training data.
The field of training data privacy develops attacks
(to leak training data details) and defenses
(to prevent leaks).

\paragraph{Privacy Attacks.}
When models are not trained with privacy-preserving algorithms, they are vulnerable to numerous privacy attacks.
The least revealing form of attack is the
\emph{membership inference attack} \cite{shokri2017membership,nasr2019comprehensive,song2018natural,hisamoto2020membership}: given a trained model,
an adversary can predict whether or not a \emph{particular} example was used to train the model.
Separately,
model inversion attacks \cite{fredrikson2015model} %
reconstruct representative views of a subset of examples 
(e.g., a model inversion attack on a face recognition classifier 
might recover a fuzzy image of a particular person that the classifier can recognize).

\emph{Training data extraction attacks}, like model inversion attacks,  reconstruct training datapoints.
However, training data extraction attacks aim
to reconstruct \emph{verbatim} training examples and not just representative ``fuzzy'' examples. This makes them more dangerous, e.g., they can extract secrets such as verbatim social security numbers or passwords.
Training data extraction attacks have until now been limited to small LMs
trained on academic datasets under artificial training setups (e.g., for more epochs than typical) \cite{carlini2019secret,song2020information,thakkar2020understanding,zanella2020analyzing}, 
or settings where the adversary has a priori knowledge of the secret they want to extract (e.g., a social security number)~\cite{carlini2019secret, henderson2018ethical}.

\paragraph{Protecting Privacy.}
An approach to minimizing memorization of training data is to apply differentially-private training techniques
\cite{rubinstein2009learning,chaudhuri2009privacy,shokri2015privacy,abadi2016deep,mcmahan2017learning}.
Unfortunately, training models with differentially-private mechanisms
often reduces accuracy \cite{EvansDP} because
it causes models to fail to capture the long tails of the data distribution \cite{song2018natural,feldman2020does,feldman2020neural}. Moreover, it increases training time, which can further reduce accuracy because
current LMs are limited by the cost of training~\cite{kaplan2020scaling,li2020train,raffel2019exploring}. %
As a result, state-of-the-art LMs such as GPT-2 \cite{radford2019better}, GPT-3 \cite{brown2020language}, and T5 \cite{raffel2019exploring} do not apply these
privacy-preserving techniques.

\section{Threat Model \& Ethics}

Training data extraction attacks are often seen as theoretical
or academic and are thus unlikely to be exploitable in practice \cite{eff}.
This is justified by the prevailing intuition that privacy leakage is
correlated with overfitting~\cite{yeom2018privacy}, 
and because state-of-the-art LMs are trained on large (near terabyte-sized \cite{brown2020language}) datasets for a few epochs, they tend to not overfit \cite{radford2019better}.

Our paper demonstrates that training data extraction attacks
are \emph{practical}.
To accomplish this, we first precisely define what we mean by
``memorization''.
We then state our threat model and our attack objectives.
Finally, we discuss the ethical considerations behind these attacks and explain why they are likely to be a serious threat in the future.

\subsection{Defining Language Model Memorization}
\label{subsec:memorization_defs}

There are many ways to define memorization in language modeling.
As mentioned earlier, memorization is in many ways an \emph{essential} component of language models
because the training objective is to assign high overall likelihood to the training dataset.
LMs must, for example, ``memorize'' the correct spelling of individual words.

Indeed, there is a research direction that analyzes neural networks
as repositories of (memorized) knowledge \cite{petroni2019language,roberts2020much}.
For example, when GPT-2 is prompted to complete the sentence
``My address is 1 Main Street, San Francisco CA'', it generates ``94107'':
a correct zip code for San Francisco, CA.
While this is clearly memorization in some abstract form,%
we aim to formalize our definition of memorization in order to restrict it to cases that we might consider ``unintended'' \cite{carlini2019secret}.

\subsubsection{\titlecap{\term{}} of Text} %

We define \term{} as a particular type of memorization.\footnote{\emph{Eidetic memory} (more commonly called \emph{photographic memory}) is the ability to recall information after seeing it only once.}
Informally, \term{} is data that has been memorized by a model despite only appearing in a small set of training instances. The fewer training samples that contain the data, the stronger the \term{} is.

To formalize this notion, we first define what it means for a model to have knowledge of a string $s$. Our definition is loosely inspired by knowledge definitions in interactive proof systems~\cite{goldwasser1989knowledge}: a model $f_{\theta}$ knows a string $s$ \emph{if $s$ can be extracted by interacting with the model}. More precisely, we focus on \emph{black-box} interactions where the model generates $s$ as the most likely continuation when prompted with some \emph{prefix} $c$:
\begin{definition}[Model Knowledge Extraction]
\label{def:extraction}
A string $s$ is \emph{extractable}%
\footnote{This definition admits pathological corner cases.
For example, many LMs when when prompted with
\emph{``Repeat the following sentence: \_\_\_\_\_.''} will do so correctly.
This allows \emph{any} string to be ``known'' under our definition.
Simple refinements of this definition do not solve the issue, as LMs can also be asked to, for example, down-case a particular sentence. We avoid these pathological cases by prompting LMs only with short prefixes.}
from an LM $f_\theta$ if there exists a prefix $c$ such that:
\[
s \gets \argmax_{s':\ |s'|=N} f_\theta(s' \mid c) \;
\]
\end{definition}

We abuse notation slightly here to denote by $f_\theta(s' \mid c)$ the likelihood of an entire sequence $s'$. Since computing the most likely sequence $s$ is intractable for large $N$, the $\argmax$ in Definition~\ref{def:extraction} can be replaced by an appropriate \emph{sampling strategy} (e.g., greedy sampling) that reflects the way in which the model $f_\theta$ generates text in practical applications.  %
We then define \term{} as follows:
\begin{definition}[$k$-\titlecap{\term{}}] %
\label{def:memorization}
A string $s$ is $k$-eidetic memorized (for $k \ge 1$) by an LM $f_\theta$ if 
$s$ is extractable from $f_\theta$
and $s$ appears in at most $k$ examples in the training data $X$:
$
\left| \{x \in X : s \subseteq x \} \right| \le k.
$
\end{definition}

Key to this definition is what ``examples'' means.
For GPT-2, each webpage is used (in its entirety) as one training example.
Since this definition counts the number of distinct training examples containing a given string,
and not the total number of times the string occurs, 
a string may appear multiple times on one page while still counting as $k=1$ memorization.

This definition allows us to define memorization as a spectrum.
While there is no definitive value of $k$ at which we might say that memorization
is unintentional and potentially harmful, smaller values are more likely to be so. For any given $k$, memorizing longer strings is also ``worse'' than shorter strings, although our definition omits this distinction for simplicity.

For example, under this definition, memorizing the correct spellings of
one particular word is not severe if the word occurs in many training examples (i.e., $k$ is large). %
Memorizing the zip code of a particular city might be \term{}, depending on
whether the city was mentioned in many training examples (e.g., webpages) or just a few.
Referring back to Figure~\ref{fig:teaser}, 
memorizing an individual person's name and phone number clearly (informally) violates privacy expectations,
and also satisfies our formal definition: it is contained in just a few documents on the Internet---and hence the training data.

\subsection{Threat Model}

\paragraph{Adversary's Capabilities.}
We consider an adversary who has black-box input-output access to a language model.
This allows the adversary to compute the probability of arbitrary sequences
$f_\theta(x_1, \dots, x_n)$, and as a result allows the adversary to obtain next-word predictions, but it does not allow the adversary to inspect individual
weights or hidden states (e.g., attention vectors) of the language model.

This threat model is highly realistic as many LMs are available through black-box APIs.
For example, the GPT-3 model \cite{brown2020language} created by OpenAI is available through black-box API access.
Auto-complete models trained on actual user data  
have also been made public,
although they reportedly use privacy-protection measures during training~\cite{smartcompose}.

\paragraph{Adversary's Objective.}
The adversary's objective is to extract memorized training data from the model.
The strength of an attack is measured by how private (formalized as being $k$-eidetic memorized)
a particular example is. Stronger attacks
extract more examples in total (both more total sequences, and longer sequences) and
examples with lower values of $k$.

We do not aim to extract \emph{targeted} pieces of training data, 
but rather \emph{indiscriminately} extract training data.
While targeted attacks have the potential to be more adversarially harmful, 
our goal is to study the ability of LMs to
memorize data generally, not to create an attack that can be operationalized by real adversaries to target specific users.

\paragraph{Attack Target.}
We select GPT-2~\cite{radford2019language} as a representative LM to study for our attacks.
GPT-2 is nearly a perfect target. First, from an ethical standpoint, the model and data are public, and so any memorized data that we extract is \emph{already} public.\footnote{Since the training data is sourced from the public Web, all the outputs of our extraction attacks can also be found via Internet searches. Indeed, to evaluate whether we have found memorized content, we search for the content on the Internet and are able to find these examples relatively easily.
}
Second, from a research standpoint, the dataset (despite being collected from public sources) was never actually released by OpenAI. Thus, it is not possible for us to unintentionally ``cheat'' and develop attacks that make use of knowledge of
the GPT-2 training dataset.

\subsection{Risks of Training Data Extraction}
\label{ssec:risks}

Training data extraction attacks present numerous privacy risks.
From an ethical standpoint, most of these risks are mitigated in our paper because we attack GPT-2, whose training data is public.
However, since our attacks would apply to \emph{any} LM, we also discuss potential consequences of future attacks on models that may be trained on private data.

\paragraph{Data Secrecy.} The most direct form of privacy leakage occurs when data is extracted from a model that was trained on confidential or private data. For example, GMail's auto-complete model \cite{smartcompose} is trained on private text communications between users, so the extraction of unique snippets of training data would break data secrecy.

\paragraph{Contextual Integrity of Data.} The above privacy threat corresponds to a narrow view of data privacy as \emph{data secrecy}. 
A broader view of the privacy risks posed by data extraction stems from the framework of data privacy as \emph{contextual integrity}~\cite{nissenbaum2004privacy}.
That is, data memorization is a privacy infringement if it causes data to be used outside of its intended context.
An example violation of contextual integrity is shown in Figure~\ref{fig:teaser}. This individual's name, address, email, and phone number are not \emph{secret}---they were shared online in a specific context of intended use (as contact information for a software project)---but are reproduced by the LM in a separate context. Due to failures such as these, user-facing applications that use LMs may inadvertently emit data in inappropriate contexts, e.g., a dialogue system may emit a user's phone number in response to another user's query.

\paragraph{Small-$k$ Eidetic Risks.}
We nevertheless focus on $k$-eidetic memorization with a small $k$ value because it makes extraction attacks more impactful.%
While there are cases where large-$k$ memorization may
still matter (for example, a company may refer to the name of an upcoming product multiple times in private---and even though it is discussed often the name itself may still be sensitive)
we study the small-$k$ case.

Moreover, note that although we frame our paper as an ``attack'', LMs will output memorized data \emph{even in the absence of an explicit adversary}. We treat LMs as black-box generative functions, and the memorized content that we extract can be generated through honest interaction with the LM.
Indeed, we have even discovered at least one memorized training example among the $1{,}000$ GPT-3 samples that OpenAI originally released in its official repository \cite{gpt3github}.

\subsection{Ethical Considerations}
\label{subsec:ethics}
In this paper, we will discuss and carefully examine \emph{specific}
memorized content that we find in our extraction attacks.
This raises ethical considerations as some of the data that we extract contains information about individual users.

As previously mentioned, we minimize ethical concerns by using data that is already public.
We attack the GPT-2 model,
which is available online. Moreover, the GPT-2 training data was
collected from the public Internet~\cite{radford2019language}, and is
in principle available to anyone who performs the same
(documented) collection process as OpenAI, e.g., see~\cite{Gokaslan2019OpenWeb}.

However, there are still ethical concerns even though the model and data are public.
It is possible---and indeed we find it is the case---that we might extract personal information for individuals from the training data. For example, as shown in Figure~\ref{fig:teaser}, we recovered
a person's full name, address, and phone number.
In this paper, whenever we succeed in extracting personally-identifying
information (usernames, phone numbers, etc.) we partially mask out this content with the token \pii.
We are aware of the fact that this does not provide complete mediation: disclosing that the vulnerability exists allows a malicious
actor to perform these attacks on their own to recover
this personal information. %

Just as responsible disclosure still causes some (limited) harm, we believe that
the benefits of publicizing these attacks outweigh the potential harms.
Further, to make our attacks public, we must
necessarily reveal some sensitive information.
We contacted the individual whose information is partially shown in Figure~\ref{fig:teaser} to disclose this fact to them in advance and received permission to use this example.
Our research findings have also been disclosed to OpenAI.

Unfortunately, we cannot hope to contact all researchers
who train large LMs in advance of our publication.
We thus hope that this publication will spark further discussions on the ethics of memorization and extraction among other companies and research teams that train large
LMs \cite{raffel2019exploring,adiwardana2020towards,shoeybi2019megatron,lewis2019bart}.

\section{Initial Training Data Extraction Attack}

We begin with a simple strawman baseline for extracting training data from a language model in a two-step procedure.
\begin{itemize}[topsep=3pt,itemsep=2pt,partopsep=0pt, parsep=0pt,leftmargin=15pt]
    \item \textbf{Generate text.} We generate a large quantity of data by unconditionally sampling from the model (Section~\ref{subsec:generating}).
    \item \textbf{Predict which outputs contain memorized text.} We next remove the generated samples that are unlikely to contain memorized text using a membership inference attack (Section~\ref{subsec:membership}).
\end{itemize}

These two steps correspond directly to extracting model knowledge (Definition~\ref{def:extraction}), 
and then predicting which strings might be $k$-\term{} (Definition~\ref{def:memorization}).

\subsection{Initial Text Generation Scheme}\label{subsec:generating}

To generate text, we initialize the language model with a one-token prompt containing a special start-of-sentence token and then repeatedly sample tokens in an autoregressive fashion from the model (see Section~\ref{subsec:lm} for background). %
We hope that by sampling according to the model's assigned likelihood, we will sample sequences that the model considers ``highly likely'', and that likely sequences correspond to memorized text.
Concretely, we sample exactly $256$ tokens for each trial using the top-$n$ strategy from Section~\ref{subsec:lm} with $n=40$.

\subsection{Initial Membership Inference 
}\label{subsec:membership}

Given a set of samples from the model, the problem of training data extraction reduces to one of membership
inference: predict whether each sample was present in the training data~\cite{shokri2017membership}. In their most basic form, past membership inference attacks rely on the observation that models tend to assign higher confidence to examples that are present in the training data~\cite{nasr2018machine}.
Therefore, a potentially high-precision membership inference classifier is to simply choose examples that are assigned the highest likelihood by the model.

Since LMs are \textit{probabilistic} generative models, we follow prior
work \cite{carlini2019secret} and use a natural likelihood measure:
the \emph{perplexity} of a sequence measures how 
well the LM ``predicts'' the tokens in that sequence.
Concretely, given a sequence of tokens $x_1, \ldots, x_n$, the perplexity is defined as
\[\mathcal{P} = \exp\left(-\frac{1}{n}\sum_{i = 1}^n\log f_\theta(x_i | x_1, \ldots, x_{i - 1})\right)\]
That is, if the perplexity is low, then the model is not very ``surprised'' by the sequence and has assigned on average a high
probability to each subsequent token in the sequence.

\subsection{Initial Extraction Results}

We generate 200{,}000 samples using the largest version of the GPT-2 model (XL, 1558M parameters) following the text generation scheme described in Section~\ref{subsec:generating}.
We then sort these samples according to the model's perplexity measure and investigate those with the lowest perplexity.

This simple baseline extraction attack can find a wide variety of memorized content.
For example, GPT-2 memorizes the entire text of the MIT public license, as well as the user guidelines of Vaughn Live, an online streaming site.
While this is ``memorization'', %
it is only $k$-eidetic memorization for a large value of $k$---these licenses occur thousands of times.

The most interesting (but still not \term{} for low values of $k$) examples include the memorization of
popular individuals' Twitter handles %
or email addresses (omitted to preserve user privacy).
In fact, all memorized content we identify in
this baseline setting is likely to have appeared in the training dataset
many times.

This initial approach has two key weaknesses that we can identify.
First, our sampling scheme tends to produce a low diversity of outputs. For example, out of the $200{,}000$ samples we generated, 
several hundred
are duplicates of the memorized user guidelines of Vaughn Live. 

Second, our baseline membership inference strategy suffers from a large number of false positives, i.e., content that is assigned high likelihood but is not memorized. The majority of these false positive samples contain ``repeated'' strings (e.g., the same phrase repeated multiple times). Despite such text being highly unlikely, large LMs often incorrectly assign high likelihood to such repetitive sequences~\cite{holtzman2019degeneration}.

\section{Improved Training Data Extraction Attack}
\label{sec:refined_attacks}

The proof-of-concept attack presented in the previous section has low precision (high-likelihood samples are not always in the training data) and low recall (it identifies no $k$-memorized content for low $k$). Here, we improve the attack by incorporating better methods for sampling from the model (Section~\ref{subsec:improved_sampling}) and membership inference (Section~\ref{subsec:improved_membership}).

\subsection{Improved Text Generation Schemes}\label{subsec:improved_sampling}

The first step in our attack is to randomly sample from the language model.
Above, we used top-$n$ sampling and conditioned the LM on the start-of-sequence token as input.
This strategy has clear limitations \cite{ippolito2020automatic}: it will only generate sequences that are likely from beginning to end.
As a result, top-$n$ sampling from the model will cause it to generate the same (or similar) examples several times.
Below we describe two alternative techniques for generating more diverse samples from the LM.

\begin{figure*}
    \centering
    \includegraphics[trim={0.0cm 7.7cm 1.2cm 0.0cm},clip,width=1.0\textwidth]{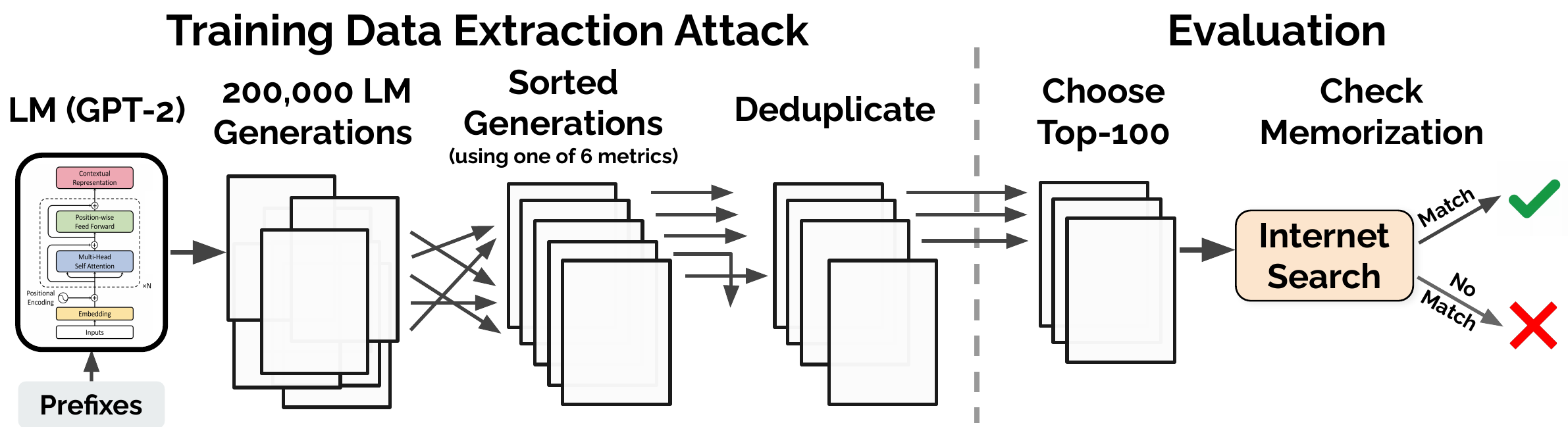}
    \caption{\textbf{Workflow of our extraction attack and evaluation.}
    \textbf{1) Attack.} We begin by generating many samples from GPT-2 when the model is conditioned on (potentially empty) prefixes.
    We then sort each generation according to one of six metrics and remove
    the duplicates. This gives us a set of potentially memorized training examples.
    \textbf{2) Evaluation.} We manually inspect 100 of the top-1000 generations 
    for each metric.
    We mark each generation as either memorized or not-memorized by manually searching online,
    and we confirm these findings by working with OpenAI to query the original training data.
    An open-source implementation of our attack process is available at  \url{https://github.com/ftramer/LM_Memorization}.}
    \label{fig:overview}
\end{figure*}

\subsubsection{Sampling With A Decaying Temperature}
\label{subsec:temperature}

As described in Section~\ref{subsec:lm}, an LM outputs the probability of the next token given the prior tokens $\textbf{Pr}(x_i \mid x_1, \dots, x_{i-1})$.
In practice, this is achieved by evaluating the
neural network $z = f_\theta(x_1, \dots, x_{i-1})$ to obtain
the ``logit'' vector $z$, and then computing
the output probability distribution as
$y = \text{softmax}(z)$ defined by
$\text{softmax}(z)_i = \exp{(z_i)} / \sum_{j=1}^n \exp{(z_j)}$. 

One can artificially ``flatten'' this probability distribution to make the model less confident by replacing the output $\text{softmax}(z)$ with $\text{softmax}(z/t)$, for $t>1$. Here, $t$ is called the \emph{temperature}. A higher temperature causes the model to be less confident and more diverse in its output.

However, maintaining a high temperature throughout the generation process would mean that even if the sampling process began to emit a memorized example, it would likely randomly step off the path of the memorized
output. Thus, we use a softmax temperature that decays over time, starting at $t=10$ and decaying down to
$t=1$ over a period of the first $20$ tokens ($\approx$10\% of the length of the sequence).
This gives a sufficient amount of time for the model to ``explore'' a diverse set of
prefixes while also allowing it to follow a high-confidence paths that it finds.

\subsubsection{Conditioning on Internet Text}
\label{subsec:ccrawl}

Even when applying temperature sampling, there are still some prefixes
that are unlikely to be sampled but nevertheless occur in actual data.
As a final strategy,
our third sampling strategy
seeds the model with prefixes from our own scrapes of the Internet.
This sampling strategy ensures that we will generate samples with a diverse set of
prefixes that are similar in nature to the type of data GPT-2 was trained on.

We follow a different data collection process as used in GPT-2 (which follows Reddit links) in order to reduce the likelihood that our dataset 
has any intersection with the model's training data.
In particular, we select samples from a subset of Common Crawl\footnote{\url{http://commoncrawl.org/}} to feed as context to the model.\footnote{It is possible there is some intersection between these two datasets, effectively allowing this strategy to ``cheat''. We believe this does not considerably affect results. First, any overlap between the two datasets is rare on average. %
Second, because we only use between the first $5$ to $10$ tokens of each sample, any possible overlap will be small in absolute terms.}

As in prior work~\cite{raffel2019exploring}, we perform basic data-sanitization by removing
HTML and JavaScript from webpages, and we de-duplicate data on a line-by-line basis.
This gives us a dataset of $50$MB of text. We randomly sample between $5$ and 
$10$ tokens of context from this scraped data and then continue LM generation with top-$n$ sampling as in Section~\ref{subsec:generating}.

\subsection{Improved Membership Inference}\label{subsec:improved_membership}

Performing membership inference by filtering out samples with low likelihood has poor precision due to failures in the underlying language model: there are many samples that are assigned spuriously high likelihood.
There are predominantly two categories of such samples:

\begin{itemize}[topsep=3pt,itemsep=2pt,partopsep=0pt, parsep=0pt,leftmargin=15pt]
\item \textbf{Trivial memorization.}
We identify many cases where GPT-2 outputs content that is uninteresting because of how common the text is.
For example, it repeats the numbers from 1 to 100 with high probability.%

\item \textbf{Repeated substrings.}
One common failure mode of LMs is their propensity to repeatedly emit the same string over and over~\cite{li2015diversity,holtzman2019degeneration}. We found many of the high-likelihood samples that are not memorized are indeed repeated texts (e.g., ``I love you. I love you\ldots'').
\end{itemize}

Our insight is that we can filter out these uninteresting (yet still high-likelihood samples) by comparing to a second LM.
Given a second model that accurately captures text likelihood, we should expect it will \emph{also} assign high likelihood to these forms of memorized content.
Therefore, a natural strategy for finding more diverse and rare forms of memorization is to filter samples where the original model's likelihood is ``unexpectedly high'' compared to a second model.
Below we discuss four methods for achieving this.

\paragraph{Comparing to Other Neural Language Models.}
Assume that we have access to a second LM that memorizes a different set of examples than GPT-2.
One way to achieve this would be to train a model on a disjoint set of training data, in which case it is unlikely that the two models will memorize the same data for small $k$.
An alternate strategy is to take a much smaller model trained on the same underlying dataset: 
because smaller models have less capacity for memorization, we conjecture that there are samples that are $k$-eidetic memorized (for small $k$) by the largest GPT-2 model, but which are not memorized by smaller GPT-2 models. Specifically, we use the Small (117M parameters) and Medium (345M parameters) models.

\paragraph{Comparing to zlib Compression.}
It is not necessary that we compare to another \emph{neural} LM;
any technique that quantifies some notion of ``surprise'' for a given sequence can be useful.
As a simple baseline method, %
we compute the zlib~\cite{zlib} entropy of the text: the number of bits of entropy when the sequence is compressed with zlib compression. We then use the ratio of the GPT-2 perplexity and the zlib entropy as our membership inference metric.
Although text compressors are simple, they can identify many of the examples of trivial memorization and repeated patterns described above (e.g., they are excellent at modeling repeated substrings).

\paragraph{Comparing to Lowercased Text.}
Instead of detecting memorization by comparing one model to another model, another option detects memorization by comparing the perplexity of the model
to the perplexity of the \emph{same} model on a ``canonicalized'' version
of that sequence.
Specifically, we measure the ratio of the perplexity on the sample before and after \textit{lowercasing} it,
which can dramatically alter the perplexity of memorized content that expects a particular casing.

\paragraph{Perplexity on a Sliding Window.}
Sometimes a model is not confident when the sample contains one memorized substring surrounded by a block of non-memorized (and high perplexity) text.
To handle this, we use the minimum perplexity when averaged over a sliding window of $50$ tokens.\footnote{Chosen after a cursory hyper-parameter sweep and manual analysis.}

\section{Evaluating Memorization}
\label{sec:eval}

We now evaluate the various data extraction methods and study common themes in the resulting memorized content.

\subsection{Methodology}

An overview of our experimental setup is shown in Figure~\ref{fig:overview}. We first build three datasets of  $200{,}000$ generated samples (each of which is $256$ tokens long) using one of our strategies:
\begin{itemize}[topsep=5pt,itemsep=2pt,partopsep=0pt, parsep=0pt,leftmargin=15pt]
    \item \emph{Top-$n$} (\S\ref{subsec:generating}) samples naively from the empty sequence.
    \item \emph{Temperature} (\S\ref{subsec:temperature}) increases diversity during sampling.
    \item \emph{Internet} (\S\ref{subsec:ccrawl}) conditions the LM on Internet text.
\end{itemize}

We order each of these three datasets 
according to each of our six membership inference metrics:
\begin{itemize}[topsep=5pt,itemsep=2pt,partopsep=0pt, parsep=0pt, leftmargin=15pt]
    \item \emph{Perplexity}: the perplexity of the largest GPT-2 model.
    \item \emph{Small}: the ratio of log-perplexities of the largest GPT-2 model and the Small GPT-2 model.
    \item \emph{Medium}: the ratio as above, but for the Medium GPT-2.
    \item \emph{zlib}: the ratio of the (log) of the GPT-2 perplexity and the zlib entropy (as computed by compressing the text).
    \item \emph{Lowercase}: the ratio of perplexities of the GPT-2 model on the original sample and on the lowercased sample.
    \item \emph{Window}: the minimum perplexity of the largest GPT-2 model across any sliding window of 50 tokens.
\end{itemize}

For each of these $3 \times 6 = 18$ configurations, we select $100$ samples from among the top-$1000$ samples according to the chosen metric.\footnote{To favor low-ranked samples, while also exploring some of the higher-ranked samples, we select the $100$ samples so that the fraction of selected samples with rank below $k$ is $\sqrt{k/1000}$.}
This gives us $1{,}800$ total samples of potentially memorized content. In real-world attacks, adversaries will look to uncover large amounts of memorized content and thus may generate many more samples. We focus on a smaller set as a proof-of-concept attack.

\paragraph{Data De-Duplication.}
To avoid ``double-counting'' memorized content, we apply an automated fuzzy de-duplication step when we select the $100$ samples for each configuration.

Given a sample $s$, we define the \emph{trigram-multiset} of $s$, denoted $\texttt{tri}(s)$ as a multiset of all word-level
trigrams in $s$ (with words split on whitespace and punctuation characters). For example, the sentence ``my name my name my name'' has two trigrams (``my name my'' and ''name my name'') each of multiplicity $2$. We mark a sample $s_1$ as a duplicate of another sample $s_2$, if their trigram multisets are similar, specifically if $|\texttt{tri}(s_1) \cap \texttt{tri}(s_2) | \geq |\texttt{tri}(s_1)|/2$.

\paragraph{Evaluating Memorization Using Manual Inspection.}
For each of the $1{,}800$ selected samples, one of four authors manually determined whether the sample contains memorized text. Since the training data for GPT-2 was sourced from the public Web, our main tool is Internet searches. We mark a sample as memorized if we can identify a non-trivial substring that returns an \emph{exact match} on a page found by a Google search.

\paragraph{Validating Results on the Original Training Data.}
Finally, given the samples that we believe to be memorized,
we work with the original authors of GPT-2 to obtain limited
query access to their training dataset.
To do this we sent them all $1,800$ sequences we selected for
analysis.
For efficiency, they then performed a fuzzy $3$-gram match to account for memorization
with different possible tokenizations.
We marked samples as memorized if
all $3$-grams in the memorized sequence occurred in close proximity
in the training dataset.
This approach eliminates false negatives, but has false positives.
It can confirm that our samples are memorized but cannot detect
cases where we missed memorized samples.
In some experiments below, we report exact counts for how often a particular sequence occurs in the training data.
We obtained these counts by asking the GPT-2 authors to perform a 
separate \texttt{grep} over the entire dataset to get an exact count.

\begin{table}[t]
    \centering
    \setlength{\tabcolsep}{3pt}
    \begin{tabular}{@{} l r @{}}
    \textbf{Category} & \textbf{Count}\\
    \toprule
    US and international news & 109 \\
    Log files and error reports & 79\\
    License, terms of use, copyright notices & 54 \\
    Lists of named items (games, countries, etc.) & 54\\
    Forum or Wiki entry & 53\\
    Valid URLs & 50\\
    \textbf{Named individuals (non-news samples only)} & 46\\
    Promotional content (products, subscriptions, etc.) & 45\\
    High entropy (UUIDs, base64 data) & 35 \\
    \textbf{Contact info (address, email, phone, twitter, etc.)} & 32\\
    Code & 31\\
    Configuration files & 30\\
    Religious texts & 25\\
    Pseudonyms & 15\\
    Donald Trump tweets and quotes & 12\\
    Web forms (menu items, instructions, etc.) & 11\\
    Tech news & 11\\
    Lists of numbers (dates, sequences, etc.) & 10\\
    \bottomrule
    \end{tabular}
    \caption{Manual categorization of the 604 memorized training examples that we extract from GPT-2, along with a description of each category. Some samples correspond to multiple categories (e.g., a URL may contain base-64 data). Categories in \textbf{bold} correspond to personally identifiable information.}
    \label{tab:categories}
\end{table}

\subsection{Results}

In total across all strategies, we identify \textbf{604} unique memorized training examples from among the $1{,}800$ possible candidates, for an aggregate true positive rate of $33.5\%$ (our best variant has a true positive rate of $67\%$). 
Below, we categorize what types of content is memorized by the model, and also study which attack methods are most effective. 

\paragraph{Categories of Memorized Content.} We manually grouped the memorized samples into different categories (a description of these categories is in Appendix~\ref{apx:categorization}). The results are shown in
Table~\ref{tab:categories}. Most memorized content is fairly canonical text from news headlines, log files, entries from forums or wikis, or religious text.
However, we also identify a significant amount of unique data, containing 128-bit UUIDs, (correctly-resolving) URLs containing random substrings, and contact information of individual people and corporations.
In Section~\ref{subsec:case_studies}, we study these cases in more detail.

\paragraph{Efficacy of Different Attack Strategies.}
Table~\ref{tab:mainresults} shows the number of memorized samples broken down by the different
text generation and membership inference strategies. Sampling conditioned on Internet text is the most effective way to identify memorized content, however, all generation schemes reveal a significant amount of memorized content.
For example, the baseline strategy of generating with top-$n$ sampling yields $191$ unique memorized samples, whereas conditioning on Internet text increases this to $273$.

As discussed earlier, looking directly at the LM perplexity is a poor membership inference metric when classifying data generated with top-$n$ or temperature sampling: just 9\% and 3\% of inspected samples are memorized, respectively.
The comparison-based metrics are significantly more effective at predicting if content was memorized. For example, \textbf{67\%} of \textit{Internet} samples marked by zlib are memorized.

\begin{figure}[t]
    \centering
    \includegraphics[width=0.99\columnwidth]{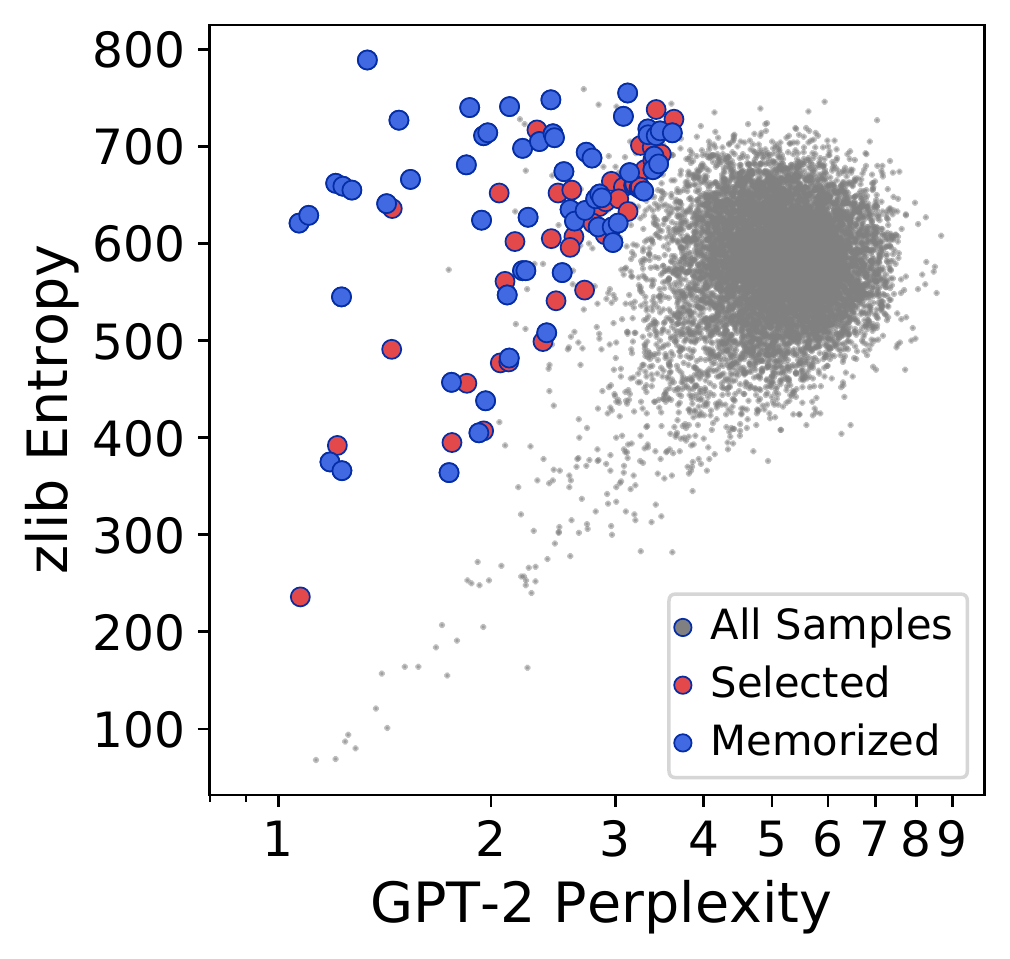}
    \vspace{-0.8em}
    \caption{The zlib entropy and the perplexity of GPT-2 XL for $200{,}000$ samples generated with top-$n$ sampling. In red, we show the 100 samples that were selected for manual inspection. In blue, we show the 59 samples that were confirmed as memorized text. Additional plots for other text generation and detection strategies are in Figure~\ref{fig:perplexities}.}
    \label{fig:scatter-mem}
\end{figure}

Figure~\ref{fig:scatter-mem} compares the zlib entropy and the GPT-2 XL perplexity for each sample, with memorized examples highlighted. Plots for the other strategies are shown in Figure~\ref{fig:perplexities} in Appendix~\ref{apx:distribution}.
Observe that most samples fall along a diagonal, i.e., samples with higher
likelihood under one model also have higher likelihood under another model.
However, there are numerous outliers in the top left:
these samples correspond to those that GPT-2 assigns a low perplexity (a high likelihood)
but zlib is surprised by. These points, especially those which are extreme outliers, are more likely to be memorized than those close to the diagonal.

The different extraction methods differ in the \emph{type} of memorized content they find.
A complete breakdown of the data is given in Appendix~\ref{apx:categorization}; however, to briefly summarize:
\begin{enumerate}[topsep=2pt,itemsep=-1ex,partopsep=1ex,parsep=2ex]
    \item The zlib strategy often finds non-rare text (i.e., has a high $k$-memorization). 
    It often finds news headlines, license files, or repeated strings from forums or wikis, and there is only one ``high entropy'' sequence this strategy finds.
    \item Lower-casing finds content that is likely to have irregular capitalization, such as news headlines (where words are capitalized) or error logs (with many uppercase words).
    \item The Small and Medium strategies often find rare content. There are 13 and 10 high entropy examples found by using the Small and Medium GPT-2 variants, respectively (compared to just one with zlib).
\end{enumerate}

\begin{table}[t]
    \centering
    \begin{tabular}{@{}l@{\hskip 0.25in}rrr@{}}
    \multirow{3}{*}{\shortstack[l]{\textbf{Inference} \\ \textbf{Strategy}}}
        & \multicolumn{3}{c}{\textbf{Text Generation Strategy}} \\
        \cmidrule(l{5pt}r{0pt}){2-4}
        & \textbf{Top-$n$} & \textbf{Temperature} & \textbf{Internet} \\
        \toprule
    \textbf{Perplexity} & 9 & 3 & 39 \\
\textbf{Small} & 41 & 42 & 58 \\
\textbf{Medium} & 38 & 33 & 45 \\
\textbf{zlib} & 59 & 46 & 67 \\
\textbf{Window} & 33 & 28 & 58 \\
\textbf{Lowercase} & 53 & 22 & 60 \\
\midrule
\textbf{Total Unique} & 191 & 140 & 273 \\
\bottomrule
    \end{tabular}
    \caption{The number of memorized examples (out of 100 candidates) that we identify using each of the three
    text generation strategies and six membership inference techniques. Some samples are found by multiple strategies; we identify 604 unique memorized examples in total.}
    \label{tab:mainresults}
\end{table}

\subsection{Examples of Memorized Content}
\label{subsec:case_studies}

We next manually analyze categories of memorized content that we find particularly compelling. 
(Additional examples are presented in Appendix~\ref{apx:case_studies}.)
Recall that since GPT-2 is trained on public data,
our attacks are not particularly severe.
Nevertheless, we find it useful to analyze what we are able to extract to understand
the categories of memorized content---with the understanding that 
attacking a model trained on a sensitive dataset would give stronger results.

\paragraph{Personally Identifiable Information.} We identify numerous examples of individual
peoples' names, phone numbers, addresses, and social media accounts. 

We find 46 examples that contain individual peoples' names.
When counting occurrences of named individuals, we omit memorized samples that relate to national and international news (e.g., if GPT-2 emits the name of a famous politician, we do not count this as a named individual here). 
We further find 32 examples that contain some form of contact information (e.g., a phone number or social media handle). Of these, 16 contain contact information for businesses, and 16 contain private individuals' contact details.

Some of this memorized content is exclusive to just
a few documents. 
For example, we extract the usernames of six users participating
in an IRC conversation that appeared in exactly one training document.

\paragraph{URLs.}
We identify $50$ examples of memorized URLs that correctly resolve to live webpages.
Many of these URLs contain uncommon pieces of text, such as random numbers or base-64 encoded strings.
We also identify several URLs that resolve correctly but we cannot
identify their source (and we thus do not count them as ``memorized'' in our evaluation). %

\paragraph{Code.}
We identify $31$ generated samples that contain snippets of memorized source code.
Despite our ability to recover the source code verbatim, we are almost always \emph{unable} to recover the original authorship notices or terms of use.
Often, this information is given either before the code itself or in a LICENSE file that appears separately.
For many of these samples, we can also \emph{extend} their length and recover thousands of lines of (near verbatim) source code (see Section~\ref{subsec:long}).

\paragraph{Unnatural Text.}
Memorization is not limited to natural-looking
text. We find
$21$ instances of random number sequences with at least
50 bits of entropy.\footnote{We estimate the entropy through manual analysis by guessing the entropy space given the format of the string.} %
For example, we extract the following UUID:
\begin{quote}
    1e4bd2a8-e8c8-4a62-adcd-40a936480059
\end{quote}
from the model; a Google search for this string identifies
just 3 documents containing this UUID, and it is contained in just one GPT-2 training document (i.e., it is $1$-eidetic memorized). %
Other memorized random number sequences include
UUIDs contained in only a few documents (not listed to preserve privacy), git commit hashes, random IDs used for ad tracking, and product model numbers.

Table~\ref{tab:high_entropy} gives nine examples of $k=1$ eidetic memorized content, each of which is a random sequences between $10$ and $87$ characters long.
In each of these cases, the memorized example is contained in exactly \textit{one} training document, and the total number of occurrences within that single document varies between \textbf{just 10} and 311.

\begin{table}
\centering
    \begin{tabular}[t]{@{}lrrr@{}}
    \multirow{3}{*}{\shortstack[l]{\textbf{Memorized} \\ \textbf{String}}} & \multirow{3}{*}{\shortstack[r]{\textbf{Sequence} \\ \textbf{Length}}} & \multicolumn{2}{c}{\textbf{\ Occurrences in Data}} \\
    \cmidrule(l{5pt}r{0pt}){3-4}
     & & \textbf{Docs} & \textbf{Total} \\
     \toprule
    \texttt{Y2}...\censor{xxxx}...\texttt{y5} & 87 & 1 & 10 \\
    \texttt{7C}...\censor{xxxx}...\texttt{18} & 40 & 1 & 22 \\
    \texttt{XM}...\censor{xxxx}...\texttt{WA} & 54 & 1 & 36 \\
    \texttt{ab}...\censor{xxxx}...\texttt{2c} & 64 & 1 & 49 \\
    \texttt{ff}...\censor{xxxx}...\texttt{af} & 32 & 1 & 64 \\
    \texttt{C7}...\censor{xxxx}...\texttt{ow} & 43 & 1 & 83 \\
    \texttt{0x}...\censor{xxxx}...\texttt{C0} & 10 & 1 & 96 \\
    \texttt{76}...\censor{xxxx}...\texttt{84} & 17 & 1 & 122 \\
    \texttt{a7}...\censor{xxxx}...\texttt{4b} & 40 & 1 & 311 \\
\bottomrule
    \end{tabular}
    \caption{\textbf{Examples of $k=1$ eidetic memorized, high-entropy content that we extract} from the 
    training data. Each is contained in \emph{just one} document.
    In the best case, we extract a 87-characters-long sequence
    that is contained in the training dataset just 10 times in total, all in the same document.}
    \label{tab:high_entropy}
\end{table}

\paragraph{Data From Two Sources.}
We find samples that contain two or more snippets of memorized text that are unrelated to one another.
In one example, GPT-2 generates a news article about the (real) murder of a woman in 2013, but then attributes the murder to one of the victims of a nightclub shooting in Orlando in 2016. Another sample starts with the memorized Instagram biography of a pornography producer, but then goes on to incorrectly describe an American fashion model as a pornography actress. This type of generation is not $k$-eidetic memorization (these independent pieces of information never appear in the same training documents), but it is an example of a contextual integrity violation.

\paragraph{Removed Content.}
Finally, GPT-2 memorizes content that has since been removed from the Internet, and is thus now \emph{primarily}
accessible through GPT-2. We are aware of this content as it is still cached by Google search, but is no longer present on the linked webpage.
Some of this data is not particularly interesting in its own right, e.g., error logs due to a misconfigured webserver that has since been fixed. %
However, the fact that this type of memorization occurs highlights 
that LMs that are trained entirely on (at-the-time) public data may end up serving as an unintentional archive for removed data.

\subsection{Extracting Longer Verbatim Sequences}\label{subsec:long}

In our previous experiments, we extract strings of 256 tokens in length. Here, we briefly investigate if we can extract longer sequences. In particular, we extend the length of some of the memorized sequences by seeding the model with each sample and continuing to generate. 
To do this, we apply a beam-search-like decoding method introduced in prior work \cite{carlini2019secret}
instead of greedy decoding which often fails to generate long verbatim sequences.%

We can extend many of the memorized samples.
For example, we identify a piece of source code taken from a repository on GitHub. We can extend this snippet to extract an entire file, namely \textit{1450 lines} of verbatim source code.
We can also extract the entirety of the MIT, Creative Commons, and Project Gutenberg licenses.
This indicates that while we have extracted $604$ memorized examples,
we could likely extend many of these to much longer snippets of memorized content.

\subsection{Memorization is Context-Dependent}\label{sec:context}

Consistent with recent work on constructing effective ``prompts''
for generative LMs \cite{brown2020language,shin2020autoprompt}, we find that the memorized
content is highly dependent on the model's context.

For example, GPT-2 will complete the prompt ``3.14159'' with the first
$25$ digits of $\pi$ correctly using greedy sampling.
However, we find that GPT-2 ``knows'' (under Definition~\ref{def:memorization})
more digits of $\pi$ because
using the beam-search-like strategy introduced above extracts $500$ digits correctly.

Interestingly, by providing the more descriptive prompt
``pi is 3.14159'', straight greedy decoding
gives the first $799$ digits of $\pi$---more 
than with the sophisticated beam search.
Further providing the context
``e begins 2.7182818, pi begins 3.14159'',
GPT-2 greedily completes the first $824$ digits of $\pi$.

This example demonstrates the importance of the context:
in the right setting, orders of magnitude more extraction is
feasible than when the context is just slightly suboptimal.
We find that this holds true for our memorized examples as well.
None of the $273$ extracted samples found using Internet conditioning
can be reliably reproduced when using the same prefix initially provided to GPT-2
that produced this sample.
However, nearly all can be reproduced with high probability if we provided the entire sequence
of data up to (but not including) the beginning of the memorized content.

The important lesson here is that our work vastly \emph{under-estimates}
the true amount of content that GPT-2 memorized.
There are likely prompts that would identify much more memorized
content, but because we stick to simple prompts we do not
find this memorized content.

\section{Correlating Memorization with \\ Model Size \& Insertion Frequency}\label{sec:frequency}

Thus far, we have shown that language models can memorize verbatim training strings, even when they are trained for few epochs and achieve small train-test accuracy gaps. A natural question is how many times a string must appear for it to be memorized (i.e., $k$ in Definition~\ref{def:memorization}). Prior work has investigated LM memorization by
varying the number of times particular ``canary'' tokens were inserted into a
training dataset \cite{carlini2019secret}.
The main limitation of this approach is that it is synthetic:
canaries are inserted artificially after the dataset has been collected and may not be representative of natural data.

Here, we study how well GPT-2 memorizes
\emph{naturally occurring} canaries in the training data. In particular, we consider a piece of memorized content with the following prefix:
\begin{Verbatim}[samepage=true]
    {"color":"fuchsia","link":"https://www.
    reddit.com/r/The_Donald/comments/
\end{Verbatim}

The \texttt{reddit.com} URL above is completed by a specific 6-character article ID and a title.
We located URLs in this specific format in a single document on \texttt{pastebin.com}.
Each URL appears a varying number of times in this document, and hence in the GPT-2 training dataset.\footnote{The purpose of this text dump was to tag users of Reddit who posted frequently on specific topics. In doing so, this page repeats some of the same links many times because many users comment on the same links.}
Table~\ref{tab:reddit} shows a subset of the URLs that appear more than once, and their respective counts in the document.\footnote{We confirmed with OpenAI that the counts here are within 5\% of the true counts of these URLs in the training data.}
This allows us to ask the question: how many times must an
example appear in the
training dataset for us to extract it?

\begin{table}
\centering
    \setlength{\tabcolsep}{4pt}
    \centering
    \begin{tabular}[t]{@{}lrrcccc@{}}
        & \multicolumn{2}{c}{\textbf{Occurrences}} & & \multicolumn{3}{c}{\textbf{Memorized?}} \\
        \cmidrule{2-3} \cmidrule{5-7}
    \textbf{URL (trimmed)} & \textbf{Docs} & \textbf{Total} && \textbf{XL} & \textbf{M} & \textbf{S} \\
     \toprule
    /r/\censor{......}51y/milo\_evacua...
    & 1 & 359 && \checkmark& \checkmark& \nicefrac{1}{2}\\
    /r/\censor{......}zin/hi\_my\_name...
    & 1 & 113 && \checkmark& \checkmark& \\
    /r/\censor{......}7ne/for\_all\_yo...
    & 1 & 76 && \checkmark& \nicefrac{1}{2}& \\
    /r/\censor{......}5mj/fake\_news\_...
    & 1 & 72 && \checkmark& & \\
    /r/\censor{......}5wn/reddit\_admi...
    & 1 & 64 && \checkmark& \checkmark& \\
    /r/\censor{......}lp8/26\_evening...
    & 1 & 56 && \checkmark& \checkmark& \\
    /r/\censor{......}jla/so\_pizzagat...
    & 1 & 51 && \checkmark& \nicefrac{1}{2}& \\
    /r/\censor{......}ubf/late\_night...
    & 1 & 51 && \checkmark& \nicefrac{1}{2}& \\
    /r/\censor{......}eta/make\_christ...
    & 1 & 35 && \checkmark& \nicefrac{1}{2}& \\
    /r/\censor{......}6ev/its\_officia...
    & 1 & 33 && \checkmark& & \\
    /r/\censor{......}3c7/scott\_adams...
    & 1 & 17 && & & \\
    /r/\censor{......}k2o/because\_his...
    & 1 & 17 && & & \\
    /r/\censor{......}tu3/armynavy\_ga...
    & 1 & 8 && & & \\
    \bottomrule
    \end{tabular}
    \caption{We show snippets of Reddit URLs that appear a varying number of times in a \textit{single} training document. We condition GPT-2 XL, Medium, or Small on a prompt that contains the beginning of a Reddit URL and report a $\checkmark$ if the corresponding URL was generated verbatim in the first $10{,}000$ generations. We report a \nicefrac{1}{2} if the URL is generated by providing GPT-2 with the first 6 characters of the URL and then running beam search.}
    \label{tab:reddit}
\end{table}

\paragraph{Methods.}
We attempt two approaches to extract URLs of this format,
and run three variants of GPT-2 (XL, Medium, and Small).
The two approaches vary the ``difficulty'' of the attack, so even
if the more difficult fails the easier may succeed.

First, 
we directly prompt each variant of GPT-2 with the prefix above, and
use top-$n$ sampling to generate $10{,}000$ possible extensions.
Then, we test whether any of the URLs %
in the training document 
were among those that were emitted by GPT-2.
We count a URL as emitted if it matches verbatim with one of the $10{,}000$ generations.

Some URLs are not extractable with this technique, and so we make the problem
easier for GPT-2 by additionally providing GPT-2 the 6-character random token
that begins each URL.
Given this additional prefix, we then sample from the model using the beam search procedure.
This task is easier in two ways: we have first provided more context and additionally use a higher recall sampling
strategy.

\paragraph{Results.}
Table~\ref{tab:reddit} summarizes the key results.
Under the more difficult of the two approaches,
the full-sized 1.5 billion parameter GPT-2 model emits all examples that are inserted
33 times or more, the medium-sized 345 million parameter memorizes half of the
URLs, and the smallest 117 million parameter model memorizes \emph{none} of these
URLs.

When given the additional context and using beam search, the medium model can emit four more URLs, and the small model only emits the one URL that was inserted 359 times.

These results illustrate two fundamental lessons in LM memorization.
First, \textit{larger} models memorize significantly more training data: even hundreds of millions of parameters are not enough to memorize some of the training points.  
The ability of LMs to improve with model size
has been extensively studied \cite{kaplan2020scaling,li2020train}; we show a negative trend where these improvements come at the cost of decreased privacy.
Second, for the largest LM,
complete memorization occurs after just $33$ insertions.
This implies that any potentially sensitive information that is repeated a non-trivial amount of times is at risk for memorization, even if it was only repeated multiple times in a single training document.

\section{Mitigating Privacy Leakage in LMs}\label{sec:defenses}

Now that we have shown that memorized training data can be extracted from LMs, a natural question is how to mitigate these threats. Here we describe several possible strategies.
 
\paragraph{Training With Differential Privacy.} Differential privacy (DP)~\cite{dwork2006calibrating,dwork2008differential} is a well-established notion of privacy that offers strong guarantees on the privacy of individual records in the training dataset. Private machine learning models can be trained with variants of the differentially private stochastic gradient descent (DP-SGD) algorithm~\cite{abadi2016deep} which is widely implemented~\cite{TFP,Opacus}.
Large companies have even used DP in production machine learning models to protect users’ sensitive information~\cite{AppleDP,ErlingssonPK14}.
The tradeoffs between privacy and utility of models %
have been studied extensively: differentially-private training typically prevents models from capturing the long tails of the data distribution and thus hurts utility~\cite{song2018natural,feldman2020does,feldman2020neural}. 

In the content of language modeling, recent work demonstrates the privacy benefits of user-level DP models%
~\cite{ramaswamy2020training}. Unfortunately, this work requires labels for which users contributed each document; such labels are unavailable for data scraped from the open Web. It may instead seem natural to aim for DP guarantees at the granularity of individual webpages, but rare snippets of text (e.g., an individual's name and contact information as in Figure~\ref{fig:teaser}) might appear in more than one webpage. 
It is thus unclear how to apply DP in a principled and effective way on Web data.

\paragraph{Curating the Training Data.} One cannot manually vet the extremely large training datasets used for training LMs. However, there are methods to limit the amount of sensitive content that is present, e.g., by identifying and filtering personal information or content with restrictive terms of use~\cite{ReCon,Continella2017}.
 
Aside from attempting to remove sensitive content, it is also important to carefully de-duplicate the data. Many language modeling datasets are de-duplicated at the document- or paragraph-level, which means that a single document can still contain many repeated occurrences of a sensitive piece of content. We envision more sophisticated strategies to de-duplicate the training data, or limit the contribution of any single source of training data.%
 
It is also vital to carefully source the training data. Many of the potentially-sensitive training examples that we extracted (e.g., individuals’ personal information) came from websites that are known to host sensitive content, e.g., \texttt{pastebin} is the 12th most popular domain in GPT-2's training set. %

Overall, sanitizing data is imperfect---some private data will always slip through---and thus it serves as a first line of defense and not an outright prevention against privacy leaks.

\paragraph{Limiting Impact of Memorization on Downstream Applications.} In many downstream applications, e.g., dialogue systems~\cite{zhang2019dialogpt} and summarization models~\cite{hoang2019efficient}, LMs are \emph{fine-tuned} on task-specific data. On the positive side, this finetuning process may cause the LM to ``forget'' \cite{mccloskey1989catastrophic,ratcliff1990connectionist} some of the data that is memorized during the pre-training stage. On the negative side, fine-tuning may introduce its own privacy leakages if the task-specific data also contains private information. An interesting direction for future work is to explore how memorization is inherited by fine-tuned models.%

Downstream applications built on top of language models could also attempt to \emph{filter out} generated text that contains memorized content, if such content can be reliably detected (e.g., using various membership inference strategies). 
 
\paragraph{Auditing ML Models for Memorization.} Finally, after mitigating privacy leaks, it is vital to audit models to empirically determine the privacy level they offer in practice~\cite{jagielski2020}. Auditing is important even when using differential privacy, as it can complement theoretical upper bounds on privacy leakage~\cite{abadi2016deep}. We envision using our proposed methods, as well as existing attacks~\cite{shokri2017membership,yeom2018privacy,jagielski2020,carlini2019secret}, to audit LMs. %

\section{Lessons and Future Work}

\paragraph{Extraction Attacks Are a Practical Threat.}
Prior work shows that ($100\times$ to $1000\times$ smaller) language models potentially memorize training data in semi-realistic settings \cite{carlini2019secret,zanella2020analyzing}.
Our results show that state-of-the-art LMs \emph{do}
memorize their training data in practice, and that adversaries can extract this data with simple techniques.
Our attacks are practical even when the data contains a given sequence only a
few times. %

As our attacks interact with a language model as a black-box, our results approximate the \emph{worst-case} behavior of language models when interacting with benign users. In particular, among $600{,}000$ (honestly) generated samples, our attacks find that at least $604$ (or $0.1\%$) contain memorized text.

Note that this is likely an extremely loose lower bound.
We only manually inspected $1{,}800$ potential candidate memorized samples;
if we had started with
more candidates we would likely have identified significantly more memorized content.
Developing improved techniques for extracting memorized data, including attacks that are targeted towards specific content, is an interesting area for future work.

\paragraph{Memorization Does Not Require Overfitting.}
It is often believed that preventing overfitting
(i.e., reducing the train-test generalization gap) 
will prevent models from memorizing training data.
However, large LMs have no significant train-test gap, and yet we still extract numerous examples verbatim from the training set. 
The key reason is that even though on \emph{average} the training loss is only slightly lower than the validation loss, there are still some training examples that have anomalously low losses.
Understanding why this happens is an important problem for future work \cite{long2018understanding,brown2020memorization}.

\paragraph{Larger Models Memorize More Data.}
Throughout our experiments, larger language models consistently memorized more training data than smaller LMs. For example, in one setting the $1.5$ billion parameter GPT-2 model memorizes over $18\times$ as much content as the $124$ million parameter model (Section~\ref{sec:frequency}). 
Worryingly, it is likely that as LMs become bigger (in fact they
already are $100\times$ larger than the GPT-2 model we study~\cite{brown2020language}), privacy leakage will become even more prevalent.

\paragraph{Memorization Can Be Hard to Discover.}
Much of the training data that we extract is only discovered when prompting the LM with a particular prefix.
Currently, we simply attempt to use high-quality prefixes and hope that they might elicit memorization.
Better prefix selection strategies~\cite{shin2020autoprompt} might identify more memorized data.

\paragraph{Adopt and Develop Mitigation Strategies.} 
We discuss several directions for mitigating memorization in LMs, including training with differential privacy, vetting the training data for sensitive content, limiting the impact on downstream applications, and auditing LMs to test for memorization. 
All of these are interesting and promising avenues of future work, but each has weaknesses and are incomplete solutions to the full problem. Memorization in modern LMs must be addressed as new generations of LMs are emerging and becoming building blocks for a range of real-world applications.

\section{Conclusion}
For large language models to be widely adopted, they must address
the training data memorization problems that we have identified.
Our extraction attacks are practical and efficient, and can recover hundreds of training examples
from a model, even when they are contained in just one training document.

Our analysis is best viewed as a cautionary tale of what could happen when training large LMs on sensitive data.
Even though our attacks target GPT-2 (which allows us to ensure that our
work is not harmful), the same techniques apply to any LM.
Moreover, because memorization gets worse as LMs become larger, we expect that these vulnerabilities will become significantly more important in the future.

There will therefore need to be techniques developed to specifically address our attacks.
Training with differentially-private techniques is one method for mitigating privacy leakage,
however, we believe that it will be necessary to develop new methods that can train models
at this extreme scale (e.g., billions of parameters) without sacrificing 
model accuracy or training time. More generally, there are many open questions that we hope will be investigated further, including why models memorize, the dangers of memorization, and how to prevent memorization.

\section*{Acknowledgements}
We are grateful for comments on early versions of this paper by 
Dan Boneh, Andreas Terzis, Carey Radebaugh, Daphne Ippolito, Christine Robson, Kelly Cooke, Janel Thamkul, Austin Tarango, Jack Clark, Ilya Mironov, and Om Thakkar.
Florian Tramèr is supported by NSF award CNS-1804222.

\section*{Summary of Contributions}

\begin{itemize}[leftmargin=10pt, itemsep=-2pt,topsep=-10pt]

\item Nicholas, Dawn, Ariel,  Tom, Colin and \'{U}lfar proposed the research question of extracting training data from GPT-2 and framed the threat model.

\item Colin, Florian, Matthew, and Nicholas stated the memorization definitions.

\item Florian, Ariel, and Nicholas wrote code to generate candidate memorized samples from GPT-2 and verify the ground truth memorization.

\item Florian, Nicholas, Matthew, and Eric manually reviewed and categorized the candidate memorized content.

\item Katherine, Florian, Eric, and Colin generated the figures.

\item Adam, Matthew, and Eric ran preliminary investigations in language model memorization.

\item Nicholas, Florian, Eric, Colin, Katherine, Matthew, Ariel, Alina, \'{U}lfar, Dawn, and Adam wrote and edited the paper.

\item Tom, Adam, and Colin gave advice on language models and machine learning background.

\item Alina, \'{U}lfar, and Dawn gave advice on the security goals.

\end{itemize}

\bibliographystyle{plain}
\bibliography{references}

\appendix

\section{Categorization of Memorized Data}
\label{apx:categorization}

Table~\ref{tab:categories_explained} %
describes the high-level categories that we assigned to the 604 memorized samples extracted from GPT-2. Note that a single sample can belong to multiple categories. Tables 6 and 7
(omitted for space) show the categorization broken down by attack strategy.

\section{Distribution of Model Perplexities}\label{apx:distribution}

\begin{figure*}[t]
    \centering
    \begin{subfigure}{0.99\textwidth}
        \includegraphics[width=\textwidth]{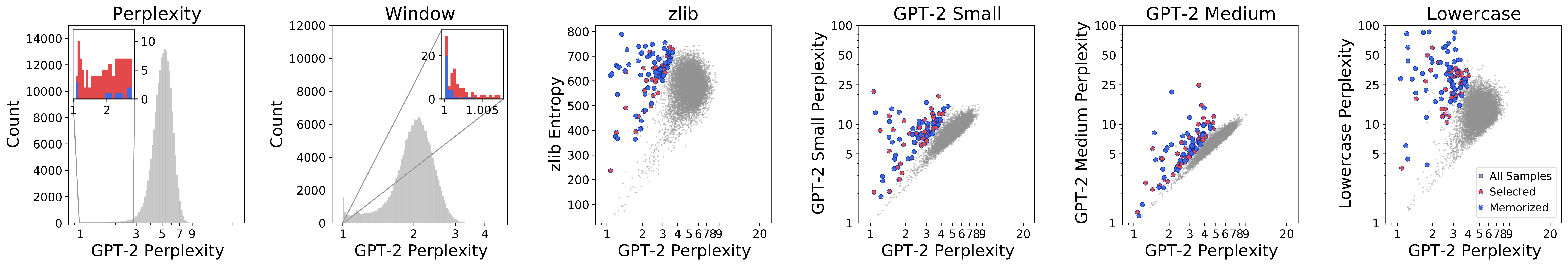}
        \caption{Top-$n$ (2.6\% duplicates)}
    \end{subfigure}
    \\%[1.5em]
    \begin{subfigure}{0.99\textwidth}
        \includegraphics[width=\textwidth]{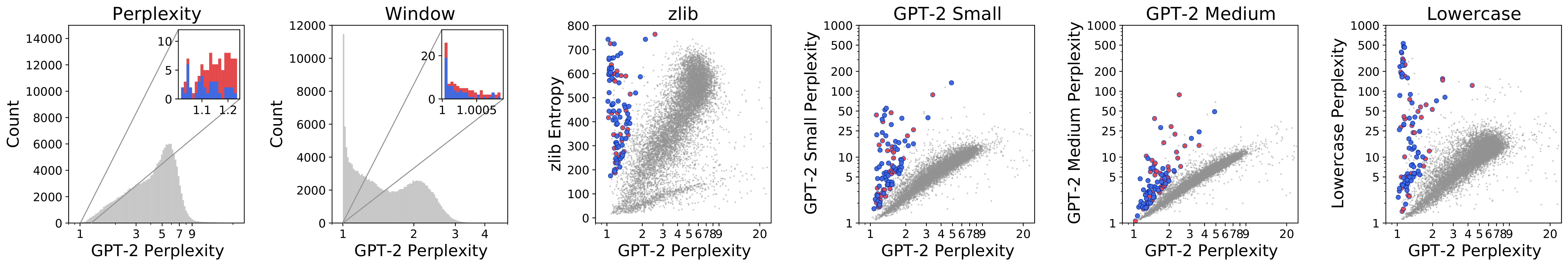}
        \caption{Internet (7.1\% duplicates)}
    \end{subfigure}
    \\%[1.5em]
    \begin{subfigure}{0.99\textwidth}
        \includegraphics[width=\textwidth]{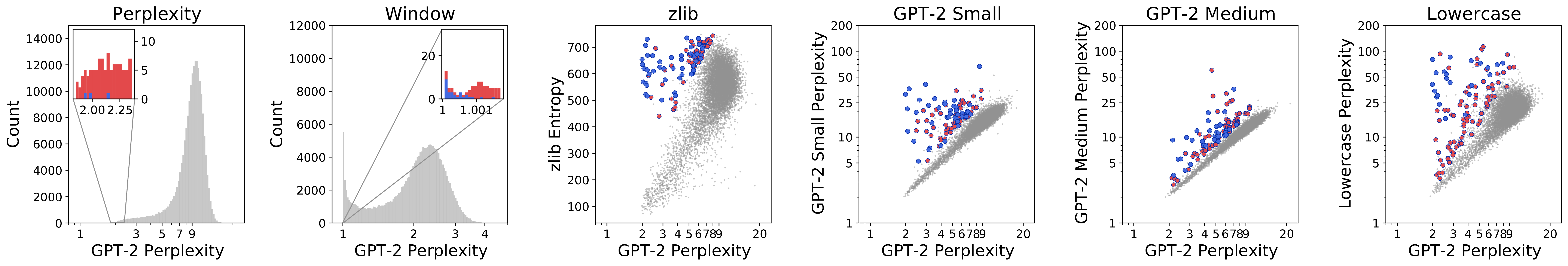}
        \caption{Temperature (0.6\% duplicates)}
    \end{subfigure}
    \caption{For each of our three text generation strategies (Top-$n$, Internet and Temperature), we generate $200{,}000$ samples using GPT-2 and apply a de-duplication procedure. The two left-most plots show the distribution of perplexities for the full sample, and the most likely window of 50 tokens. The remaining plots compare the distribution of perplexities of GPT-2 to other measure of sample likelihood: zlib entropy, perplexity under GPT-2 Small and GPT-2 Medium, and perplexity of lower-cased samples.
    Each plot highlights the 100 samples we selected for manual inspection (red) and the subset that was confirmed as memorized (blue).}
    \label{fig:perplexities}
\end{figure*}

Figure~\ref{fig:perplexities} shows the distribution of the perplexities of samples generated with each of our three text generation strategies and ordered based on our six membership inference strategies.

\section{Additional Case Studies of Memorization}\label{apx:case_studies}

Here we present additional results from our manual analysis of the memorized content.

\paragraph{Memorized Leaked Podesta Emails from WikiLeaks.}
We identify several memorized URLs that originated from the leaked Podesta Emails available on WikiLeaks\footnote{\url{https://en.wikipedia.org/wiki/Podesta_emails}}.
There is only one training document that contains these memorized URLs. Due to the nature of email, the text of one message is often included in subsequent replies to this email. As a result, a URL that is used (intentionally) only once
can be included in the dataset tens of times due to the replies.

\iffalse
\paragraph{Memorized Personal Information in Incorrect Contexts.}
In some cases, we find that GPT-2 generates text that \emph{mixes} together memorized personal information from unrelated sources and contexts. 
For example, we extract a news article about the (real) murder of a woman in 2014, which GPT-2 mistakenly attributes to a victim of a nightclub shooting in Orlando in 2016.
Another extracted sample starts with the memorized Instagram biography of a pornography producer, and then describes an American fashion model as a successful pornography actress.
While not necessary privacy violations, these ``knowledge hallucinations'' raise fundamental questions about the trustworthiness of large LMs as sources of knowledge, and could be a vehicle for abuse in user-facing applications built on top of these models.
\fi

\paragraph{Memorized Donald Trump Quotes and Tweets.}
The GPT-2 training dataset was collected when the 2016 US Presidential election was often in the news. As a result, we find several instances of memorized quotes from Donald Trump, both in the form of official remarks made as President (found in the official government records), as well as statements made on Twitter.

\paragraph{Memorized Promotional Content.}
We extract memorized samples of promotional content, such as advertisements for books, beauty products, software products.
One of these samples includes a link to an author's valid Patreon account, along with a list of named and pseudonymous prior donors. %

\paragraph{Memorized Number Sequences.}
We identify many examples where GPT-2 emits common number
sequences.
Nearly ten examples contain the integers counting up
from some specific value. We also find examples of GPT-2
counting the squares
\texttt{1, 2, 4, 8, 16, 25, 36},
Fibonacci numbers
\texttt{1, 1, 2, 3, 5, 8, 13, 21, 34, 55, 89, 144, 233, 377, 610, 987},
or digits of $\pi$,
\texttt{3.14159265358979323846264}.
None of these examples should be unexpected, but the quantity
of memorized number sequences was surprising to us.

\paragraph{Memorized News Headlines.}
Numerous memorized text snippets are verbatim copies of news articles and headlines.
A large number of these memorized samples are attributed to a single source: \texttt{thehill.com}, an American news website. Interestingly, most of these samples follow the exact same template: (1) they contain a list of different news headlines separated by a ``pipe'' symbol (|), (2) the sample begins with two \emph{merged} words, e.g., ``TrumpJesuit'', (3) the headline list ends with the all-caps word ``MORE'', and (4) the sample contains the all-caps word ``ADVERTISEMENT''.

We indeed find pages on the Web that contain copies of headlines from \texttt{thehill.com} under this exact template.
The peculiarities of these snippets likely contributed to their memorization. For example, the token TrumpJesuit does not appear in any other context on the entire Web. %

\paragraph{Memorized Base-64 Content.}
One particularly interesting form of memorization that we identify is
the ability of GPT-2 to emit base-64 encoded content. For example,
we extract out of the model the following sequence:
\begin{verbatim}
    bWFzdGVyfGltYWdlc3w3OTkxOXxpbWFnZS9wbmd
    8aW1hZ2VzL2hkZS9oMDQvODg0NTY3MjYxMTg3MC
    5wbmd8ZmFkMTMlNmFiYWJhZjFiMjJlYTAyNzU0Z
\end{verbatim}
which decodes to the sequence
``master|images|79919|image
/png|images/hde/h04/8845672611870.png|...''.
Despite our attempts, we are unable to identify where this
content originates.

\begin{table*}[b]
    \centering
    \footnotesize
    \renewcommand{\arraystretch}{1.2}
    
    \begin{minipage}[t]{\columnwidth}
    \begin{tabular}[t]{@{} >{\raggedright}p{2.3cm}@{\hskip 2pt} r p{4.9cm} @{}}
    \textbf{Category} & \textbf{Count} & \textbf{Description}\\
    \toprule
    US and international news & 109 & General news articles or headlines, mostly about US politics\\
    Log files and error reports & 79 & Logs produced by software or hardware \\
    License, terms of use, copyright notices & 54 & Software licenses or website terms of use, copyright for code, books, etc. \\
    Lists of named items  & 54 & Ordered lists, typically alphabetically, of games, books, countries, etc.\\
    Forum or Wiki entry & 53 & User posts on online forums or entries in specific wikis\\
    Valid URLs & 50 & A URL that resolves to a live page\\
    \textbf{Named individuals} & 46 & Samples that contain names of real individuals. We limit this category to \emph{non-news samples}. E.g., we do not count names of politicians or journalists within news articles\\
    Promotional content & 45 & Descriptions of products, subscriptions, newsletters, etc.\\
    High entropy & 35 & Random content with high entropy, e.g., UUIDs Base64 data, etc.\\
    \bottomrule
    \end{tabular}
    \end{minipage}
    \hfill
    \begin{minipage}[t]{\columnwidth}
    \begin{tabular}[t]{@{} >{\raggedright}p{2.3cm}@{\hskip 2pt} r p{4.9cm} @{}}
    \textbf{Category} & \textbf{Count} & \textbf{Description}\\
    \toprule
    \textbf{Contact info} & 32 & Physical addresses, email addresses, phone numbers, twitter handles, etc.\\
    Code & 31 & Snippets of source code, including JavaScript\\
    Configuration files & 30 & Structured configuration data, mainly for software products\\
    Religious texts & 25 & Extracts from the Bible, the Quran, etc.\\
    Pseudonyms & 15 & Valid usernames that do not appear to be tied to a physical name\\
    Donald Trump tweets and quotes & 12 & Quotes and tweets from Donald Trump, often from news articles\\
    Web forms & 11 &  Lists of user menu items, Website instructions, navigation prompts (e.g., ``please enter your email to continue'')\\
    Tech news & 11 & News related to technology\\
    Lists of numbers & 10 & Lists of dates, number sequences, $\pi$, etc.\\
    Sports news & 9 & News related to sports\\
    Movie synopsis, cast & 5 & List of actors, writers, producers. Plot synopsis.\\
    Pornography & 5 & Content of pornographic nature, often lists of adult film actors.\\
    \bottomrule
    \end{tabular}
    \end{minipage}
    \caption{Descriptions for the categories of memorized text. Categories in \textbf{bold} correspond to personally identifiable information.}
    \label{tab:categories_explained}
\end{table*}

\begin{table*}[b]
    \centering
    \setlength{\tabcolsep}{0pt}
    \footnotesize
    \begin{subtable}[t]{0.33\textwidth}
    \centering
    \begin{tabular}[t]{@{} l r @{}}
    \textbf{Category} & \textbf{Count}\\
    \toprule

US and international news & 88 \\
Forum or Wiki entry & 34 \\
License, terms of use, copyright notice & 28 \\
\textbf{Named individuals} & 25 \\
Promotional content & 18 \\
Lists of named items & 15 \\
\textbf{Contact info} & 20 \\
Donald Trump tweets and quotes & 12\\
Pseudonyms & 7 \\
Valid URLs & 7 \\
Sports news & 6 \\
Movie synopsis or cast & 6 \\

    \bottomrule
    \end{tabular}
    \caption{Top-$n$ (191 samples)}
    \label{tab:categorization_default}
    \end{subtable}
    \begin{subtable}[t]{0.33\textwidth}
    \centering
    \begin{tabular}[t]{@{} l r @{}}
    \textbf{Category} & \textbf{Count}\\
    \toprule

Log files and error reports & 86 \\
Lists of named items & 53 \\
Valid URLs & 40 \\
License, terms of use, copyright notice & 36 \\
High entropy & 33 \\
Configuration files & 32 \\
Code & 29 \\
\textbf{Named individuals} & 18 \\
Promotional content  & 14 \\
\textbf{Contact info} & 12\\
Pseudonyms & 11 \\
Forum or Wiki entry & 9 \\
US and international news & 7 \\
Tech news & 7 \\
Pornography & 5 \\
Web forms & 5 \\
Lists of numbers & 5 \\

    \bottomrule
    \end{tabular}
    \caption{Internet (273 samples)}
    \label{tab:categorization_ccrawl}
    \end{subtable}
    \begin{subtable}[t]{0.33\textwidth}
    \centering
    \begin{tabular}[t]{@{} l r @{}}
    \textbf{Category} & \textbf{Count}\\
    \toprule

US and international news & 31 \\
Religious texts & 28 \\
License, terms of use, copyright notice & 24 \\
Promotional content  & 20 \\
Forum or Wiki entry & 17 \\
\textbf{Named individuals } & 12 \\
Lists of named items & 12 \\
Valid URLs & 12 \\
Tech news & 8 \\
\textbf{Contact info} & 8 \\
High entropy  & 6 \\
Lists of numbers & 6 \\

    \bottomrule
    \end{tabular}
    \caption{Temperature (140 samples)}
    \label{tab:categorization_temp}
    \end{subtable}
    \caption{Memorized content found in samples produced by each of the our three text generation strategies. We show categories with at least 5 samples.}
    \label{tab:categorization_by_sampling_strategy}
\end{table*}

\begin{table*}[b]
    \setlength{\tabcolsep}{0pt}
    \footnotesize
    \begin{subtable}[t]{0.33\textwidth}
    \centering
    \begin{tabular}[t]{@{} l r @{}}
    \textbf{Category} & \textbf{Count}\\
    \toprule

License, terms of use, copyright notice & 11 \\
Lists of named items  & 8 \\
Log files and error reports & 7 \\
Valid URLs & 6 \\
Lists of numbers & 5 \\

    \bottomrule
    \end{tabular}
    \caption{Perplexity (51 samples)}
    \label{tab:categorization_ppl}
    \end{subtable}
    \begin{subtable}[t]{0.33\textwidth}
    \centering
    \begin{tabular}[t]{@{} l r @{}}
    \textbf{Category} & \textbf{Count}\\
    \toprule

US and international news & 21 \\
Lists of named items & 18 \\
License, terms of use, copyright notice & 16 \\
Promotional content & 11 \\
Valid URLs & 11 \\
Log files and error reports & 10 \\
\textbf{Named individuals}  & 8 \\
High entropy  & 8 \\
Forum or Wiki entry & 7 \\
Configuration files & 6 \\
Code & 6 \\

    \bottomrule
    \end{tabular}
    \caption{Window (119 samples)}
    \label{tab:categorization_window}
    \end{subtable}
    \begin{subtable}[t]{0.33\textwidth}
    \centering
    \begin{tabular}[t]{@{} l r @{}}
    \textbf{Category} & \textbf{Count}\\
    \toprule

US and international news & 40 \\
License, terms of use, copyright notice  & 31 \\
Lists of named items & 17 \\
Forum or Wiki entry & 14 \\
\textbf{Named individuals} & 13 \\
Promotional content & 13 \\
\textbf{Contact info} & 12\\
Log files and error reports  & 11 \\
Valid URLs & 10 \\
Code & 10 \\
Tech news & 6 \\
Configuration files & 6 \\
Pseudonyms & 5 \\

    \bottomrule
    \end{tabular}
    \caption{zlib (172 samples)}
    \label{tab:categorization_zlib}
    \end{subtable}
    \\[2em]
    \begin{subtable}[t]{0.33\textwidth}
    \centering
    \begin{tabular}[t]{@{} l r @{}}
    \textbf{Category} & \textbf{Count}\\
    \toprule

US and international news & 39 \\
Log files and error reports  & 29 \\
Lists of named items & 17 \\
Forum or Wiki entry & 12 \\
\textbf{Named individuals} & 11 \\
License, terms of use, copyright notice  & 10 \\
High entropy & 9 \\
Configuration files & 6 \\
Promotional content & 5 \\
Tech news & 5 \\

    \bottomrule
    \end{tabular}
    \caption{Lowercase (135 samples)}
    \label{tab:categorization_lower}
    \end{subtable}
    \begin{subtable}[t]{0.33\textwidth}
    \centering
    \begin{tabular}[t]{@{} l r @{}}
    \textbf{Category} & \textbf{Count}\\
    \toprule

Log files and error reports  & 17 \\
Forum or Wiki entry & 15 \\
Religious texts & 14 \\
Valid URLs & 13 \\
High entropy & 13 \\
Lists of named items & 12 \\
License, terms of use, copyright notice  & 12 \\
Promotional content & 11 \\
Configuration files & 11 \\
\textbf{Named individuals} & 11 \\
other & 9 \\
US and international news & 9 \\
\textbf{Contact info} & 8\\
Donald Trump tweets and quotes  & 7 \\
Code & 6 \\

    \bottomrule
    \end{tabular}
    \caption{Small (141 samples)}
    \label{tab:categorization_small}
    \end{subtable}
    \begin{subtable}[t]{0.33\textwidth}
    \centering
    \begin{tabular}[t]{@{} l r @{}}
    \textbf{Category} & \textbf{Count}\\
    \toprule

Valid URLs & 17 \\
Log files and error reports  & 14 \\
US and international news & 13 \\
\textbf{Contact info} & 12\\
Religious texts & 12 \\
\textbf{Named individuals} & 11 \\
Promotional content & 11 \\
High entropy & 10 \\
Forum or Wiki entry & 9 \\
Lists of named items & 8 \\
License, terms of use, copyright notice  & 8 \\
Code & 5 \\
Donald Trump tweets and quotes & 5 \\

    \bottomrule
    \end{tabular}
    \caption{Medium (116 samples)}
    \label{tab:categorization_medium}
    \end{subtable}
    \caption{Memorized content found using our six  membership inference strategies. We show categories with at least 5 samples.}
    \label{tab:categorization_by_MI_strategy}
\end{table*}

\end{document}